\documentclass[twocolumn]{aastex631} 

\definecolor{trackchange}{cmyk}{0,1,1,0}
\definecolor{explain}{cmyk}{0.93,0.67,0,0.2}

\usepackage{float}
\usepackage{bookmark} 
\bookmarksetup{numbered, open,}
\usepackage{enumitem}
\usepackage{comment}
\setlist[enumerate]{itemsep=0mm}
\usepackage{hyperref}
\usepackage{subfigure}
\usepackage{multirow}
\usepackage{xspace}
\usepackage{natbib}
\usepackage{xcolor}  
\usepackage{CJK} 
\usepackage{soul}
\usepackage{amsfonts}
\let\tablenum\undefined
\usepackage{siunitx}

\definecolor{jaredcolor}{HTML}{5D3FD3}

\newcommand{\eg}{e.g.}
\newcommand{\Msun}{M$_{\odot}$\xspace}
\newcommand{\alphaL}{$\alpha_{\Lambda}$\xspace}
\newcommand{\Lunit}{erg~s$^{-1}$\xspace}
\newcommand{\Nifs}{$^{56}$Ni\xspace}

\newcommand{\tardis}{\textsc{tardis}\xspace}
\newcommand{\kepler}{\textsc{kepler}\xspace}
\newcommand{\snec}{\textsc{snec}\xspace}
\newcommand{\flash}{\textsc{flash}\xspace}
\newcommand{\stir}{\textsc{stir}\xspace}

\newcommand{\Datalink}{\url{https://zenodo.org/records/14343229}}

\graphicspath{{./}{figures/}}

\submitjournal{ApJ}
\shorttitle{Synthetic observables of stripped high-mass star explosions}
\shortauthors{Lu et al.}

\begin{document}

\title{Physics-driven Explosions of Stripped High-Mass Stars: \\Synthetic Light Curves and Spectra of Stripped-Envelope Supernovae with Broad Lightcurves}

\correspondingauthor{Jing Lu}
\email{lujing8@msu.edu}

\author[0000-0002-3900-1452]{Jing Lu \begin{CJK*}{UTF8}{gbsn}(陆晶)\end{CJK*}}
\affil{Department of Physics and Astronomy, Michigan State University, East Lansing, MI 48824, USA}

\author[0000-0002-8825-0893]{Brandon L. Barker}
\altaffiliation{Metropolis Fellow}
\affiliation{Computational Physics and Methods, Los Alamos National Laboratory, Los Alamos, NM 87545, USA}
\affiliation{Center for Theoretical Astrophysics, Los Alamos National Laboratory, Los Alamos, NM 87545, USA}

\author[0000-0003-1012-3031]{Jared Goldberg}
\affil{Center for Computational Astrophysics, Flatiron Institute, 162 5th Avenue, New York, NY 10010, USA}

\author[0000-0002-0479-7235]{Wolfgang E. Kerzendorf}
\affil{Department of Physics and Astronomy, Michigan State University, East Lansing, MI 48824, USA}
\affil{Department of Computational Mathematics, Science, and Engineering, Michigan State University, East Lansing, MI 48824, USA}

\author[0000-0001-7132-0333]{Maryam Modjaz}
\affil{Department of Astronomy, University of Virginia, Charlottesville, VA 22904, USA}

\author[0000-0002-5080-5996]{Sean M. Couch}
\affil{Department of Physics and Astronomy, Michigan State University, East Lansing, MI 48824, USA}
\affil{Department of Computational Mathematics, Science, and Engineering, Michigan State University, East Lansing, MI 48824, USA}
\affil{Joint Institute for Nuclear Astrophysics-Center for the Evolution of the Elements, Michigan State University, East Lansing, MI 48824, USA} \affil{National Superconducting Cyclotron Laboratory, Michigan State University, East Lansing, MI 48824, USA}

\author[0000-0002-1560-5286]{Joshua V. Shields}
\affil{Department of Physics and Astronomy, Michigan State University, East Lansing, MI 48824, USA}

\author[0000-0001-7343-1678]{Andrew G. Fullard}
\affil{Department of Physics and Astronomy, Michigan State University, East Lansing, MI 48824, USA}

\begin{abstract}
Stripped-envelope supernovae (SESNe) represent a significant fraction of core-collapse supernovae, arising from massive stars that have shed their hydrogen and, in some cases, helium envelopes. 
The origins and explosion mechanisms of SESNe remain a topic of active investigation. 
In this work, we employ radiative-transfer simulations to model the light curves and spectra of a set of explosions of single, solar-metallicity, massive Wolf-Rayet (WR) stars with ejecta masses ranging from 4 to 11~\Msun, that were computed from a turbulence-aided and neutrino-driven explosion mechanism. 
We analyze these synthetic observables to explore the impact of varying ejecta mass and helium content on observable features. 
We find that the light curve shape of these progenitors with high ejecta masses is consistent with observed SESNe with broad light curves but not the peak luminosities. 
The commonly used analytic formula based on rising bolometric light curves overestimates the ejecta mass of these high-initial-mass progenitor explosions by a factor up to 2.6.
In contrast, the calibrated method by Haynie et al., which relies on late-time decay tails, reduces uncertainties to an average of 20\% within the calibrated ejecta mass range.
Spectroscopically, the \ion{He}{1} 1.083~$\mu$m line remains prominent even in models with as little as 0.02~\Msun of helium. 
However, the strength of the optical \ion{He}{1} lines is not directly proportional to the helium mass but instead depends on a complex interplay of factors such as \Nifs distribution, composition, and radiation field. 
Thus, producing realistic helium features requires detailed radiative transfer simulations for each new hydrodynamic model.
\end{abstract}



\section{Introduction}  \label{sec:intro}

Despite hydrogen being the most abundant element in the universe, $\sim$30\% of the core-collapse supernovae (CCSNe) appear hydrogen-poor or hydrogen-free \citep{Li2011, Shivvers2017}.
These objects are commonly referred to as stripped-envelope SNe (SESNe), which are the explosions of massive stars that have lost a part or all of their outer hydrogen (and helium) envelope before the core collapse \citep[\eg,][]{Wheeler1985, Clocchiatti1997, Woosley2002}. 
SESNe are comprised of three observational classes: Type IIb objects that show hydrogen only at early times in their spectra, Type Ib objects that exhibit hydrogen-free but helium-rich spectra, and Type Ic objects that lack both hydrogen and helium lines \citep{Filippenko1997, Gal-Yam2017, Modjaz2019}.
The question remains open to which extent the absence of hydrogen/helium spectra features indicates the absence of the outer hydrogen/helium-rich envelopes in the progenitor star.

It remains uncertain how exactly the outer envelope is stripped before explosion \citep[\eg,][]{Ouchi2017, Modjaz2019}.
The two major competing progenitor channels of the general SESNe population are 
1): massive single Wolf-Rayet (WR) stars that lose their outer envelope through strong and metallicity-dependent winds \citep{Conti1975, Begelman1986, Woosley1993};
and 2): intermediate-mass stars stripped by binary interactions \citep{Wheeler1985, Podsiadlowski1992, Maund2004, Eldridge2008, Yoon2010, Smith2011}.
Recent studies suggested a hybrid stripping process for SESNe progenitors, where the hydrogen is stripped by binary interactions and helium stripped through strong, metallicity-dependent stellar winds \citep{Modjaz2011, Fang2019, Sun2023}.

One important approach to inferring progenitor properties is by estimating the ejecta mass from observed light curves, as different evolutionary paths result in variations in final masses \citep[\eg,][]{Ensman1988, Hillier1991, Woosley1994, Drout2011, Sana2012}. 
However, commonly used methods for deriving ejecta mass often have large associated uncertainties \citep[\eg,][]{Wheeler2015}, emphasizing the need for self-consistent models that more accurately connect SNe properties with progenitor characteristics.

Studying ejecta composition through spectra is another way to infer the progenitor properties of SESNe.  
Recent studies indicate that progenitors with solar metallicity that undergo binary stripping produce less helium than single stars \citep{Laplace2021, Farmer2023} systematically.
However, quantifying helium in SESNe is particularly challenging, as an uncertain amount of helium can remain spectroscopically hidden \citep[\eg,][]{Harkness1987, Lucy1991, Hachinger2012, Dessart2012, Dessart2015, Williamson2021}.
Specifically, the helium spectral feature strength is strongly affected by the radiation field, which is determined by the progenitor and explosion properties. 
Thus each progenitor scenario needs a detailed study from stellar evolution to explosion dynamics and spectra formation.

For these reasons, we conduct radiative-transfer simulations to calculate the light curves and spectra of a set of solar-metallicity, massive WR stars that were evolved by \citet{Sukhbold2016} and exploded in \citet[][here after C20]{Couch2020} using a physics-driven approach. 
The turbulence-aided and neutrino-driven explosion models from C20 offer self-consistent explosion parameters, including explosion energy and remanent mass (thus the ejecta mass).
Section~\ref{sec: methods} describes the progenitors and numerical methods utilized in this work.
Section~\ref{sec: results} presents and discusses resultant light curves and spectra from the simulation framework.
Finally, we summarize our findings in Section~\ref{sec: conclusion}.

\section{Numerical Methods} \label{sec: methods}

\begin{figure*}[htb!]
\centering
\includegraphics[width=\textwidth]{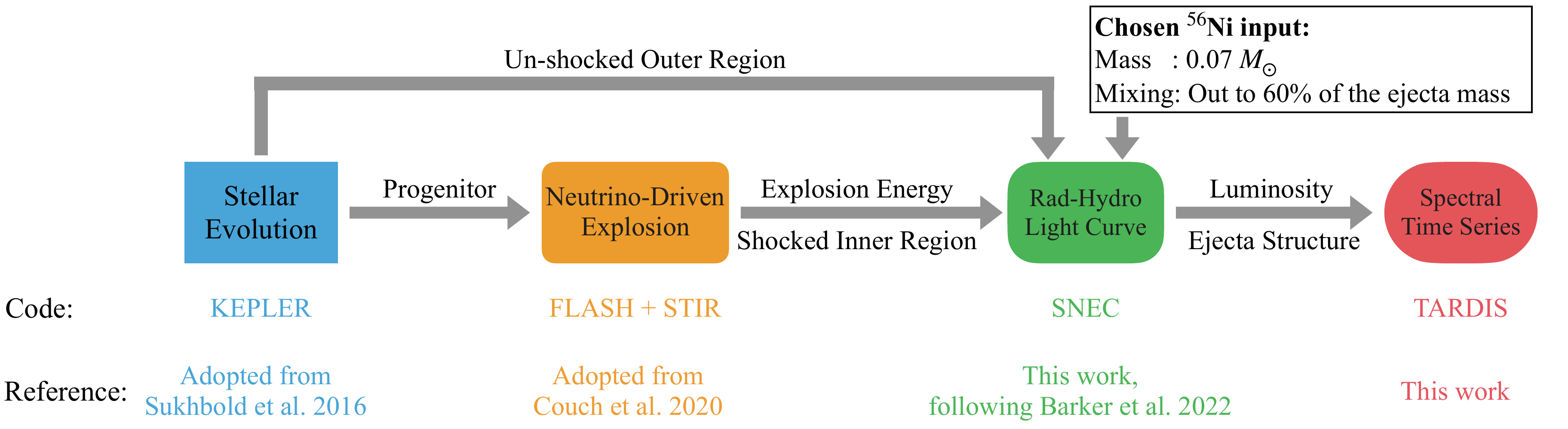}
\caption{The numerical methods flow chart of the four sequential simulation stages. The grey arrows and corresponding texts indicate the simulation inputs. The utilized code and reference for each stage are marked below. Note that all simulations are performed in 1D.}
\label{fig: method_flowchart}
\end{figure*}

We summarize the progenitors and explosions models used in this work and the radiative transfer methods for obtaining the light curves and spectra. 
In Section~\ref{subsec: Method - Progenitors}, we briefly summarize the properties of the selected massive star progenitors from \citet{Sukhbold2016} that were simulated using the \kepler code \citep{Weaver1978}. 
The neutrino-driven explosions of these progenitors are taken from \citet{Couch2020}, which utilizes the \flash \citep{Fryxell2000, Dubey2009} and \stir \citep{Couch2020} framework, as briefly discussed in Section~\ref{subsec: Method - Explosion}.
Then we use \snec \citep{Morozova2015, Morozova2016, Morozova2018} to obtain the bolometric light curves of these models, following the same manner described in \citet{Barker2022} and summarized in Section~\ref{subsec: Method - SNEC}.
Finally, the \snec outputs are mapped to \tardis \citep{Kerzendorf2014, kerzendorf_2024_13207705} to simulate the spectral time series as detailed in Section~\ref{subsec: Method - TARDIS}.
A numerical simulation series flow chart is presented in Fig.~\ref{fig: method_flowchart}.

\subsection{Progenitors} \label{subsec: Method - Progenitors}

In this work, we consider eight progenitors from \cite{Sukhbold2016} that have zero-age-main-sequence mass ($M_{\text{ZAMS}}$) ranging from 45 to 120~\Msun.
These massive solar-metalicity stars were evolved from the main sequence to the onset of core collapse using the stellar evolution code -- \kepler \citep{Weaver1978}, assuming single-star evolution without accounting for rotation or magnetic fields.
\citet{Sukhbold2016} adopted the mass loss prescription from \citet{Nieuwenhuijzen1990} for the main sequence and red giant phase, and \citet{Wellstein1999} for WR stars.
They also applied a semi-convective mixing parameter of $\alpha = 0.1$ to these massive stars. 
Details of the nucleosynthesis, mass loss, convection, and mixing treatment of these progenitor models are described in \citet[][and reference within]{Woosley2007}.

All eight of these massive progenitor stars have lost their entire H and most of the He envelopes. 
Thus, they are stripped stars, making them ideal progenitors for SESNe.
They ended as Wolf-Rayet stars before the onset of the core collapse. 
The structural properties at the time of explosion relevant for determining SNe properties are not monotonic in $M_{\text{ZAMS}}$.
The final radius ($R_{\text{pre-SN}}$), mass ($M_{\text{pre-SN}}$), and remaining helium mass ($M_{\text{He}}$) of selected progenitors right before explosion are tabulated in Table~\ref{table:pre_post_SN_prop}.

\subsection{Explosion model} \label{subsec: Method - Explosion}

The core-collapse supernova explosions of the chosen progenitors for this work were carried out in C20 as part of a larger modeling grid using all available progenitors from \cite{Sukhbold2016}.
The simulations were executed using multi-physics application code \flash \citep{Fryxell2000, Dubey2009} with \stir model \citep{Couch2020}.
C20 employed state-of-the-art neutrino transport using the two-moment ``M1'' formalism \citep{Shibata2011, Cardall2013, Oconnor2015, OconnorCouch2018}, a dense matter equation of state applicable to modeling the proto-neutron star and deep stellar interiors \citep[``SFHo,''][]{Steiner2013a}, and includes convection and turbulence effects that are consistent with high-fidelity 3D simulations. 
The \flash + \stir framework focuses on the shock revival phase due to neutrino heating in the center region of the star within seconds after the onset of core bounce. 
The details of the explosion numerical methods are described in C20.

In the \stir framework, the mixing-length parameter (\alphaL) for turbulence governs the turbulent speed and consequently affects the explodability of the star. 
C20 found a value of \alphaL $=$ 1.25 that produces consistent results with detailed 3D neutrino-driven CCSN simulations.
We adopt the models with such fiducial \alphaL value, which led to successful explosions for all eight chosen progenitor models in this work.
It is worth noting that, as long as the model successfully explodes, the explosion energy is insensitive to the choice of \alphaL among the chosen progenitor models (see Fig.~8 in C20).

However, the explosion energies of the selected models have not yet reached the asymptotic value by the end of the \stir simulation.
As stated in C20, the \stir simulation is terminated if the shock radius reaches \num{15000}~km, as the assumption of nuclear statistical equilibrium (NSE) becomes less accurate at larger radii.
We use the same approach as described in \cite{Barker2022} to estimate the asymptotic explosion energy analytically before passing it to \snec.
The average energy difference between the \stir output and estimated asymptotic value is $0.2\,\times\,10^{51}$~$erg$ in our models.

The compact remnant is considered a proto-neutron star (PNS) and, on average, has a mass of $\approx$1.8~\Msun among the selected models.
The resultant explosion properties of these models from C20 are tabulated in Table~\ref{table:pre_post_SN_prop}.

\begin{deluxetable*}{cccc|ccccc|cccc}[ht!] 
\centering
\tablecaption{Progenitor and SN properties of the models.\label{table:pre_post_SN_prop}}
\tablehead{\colhead{$M_{\text{ZAMS}}$} & \colhead{log($R_{\text{pre-SN}}$)} & \colhead{$M_{\text{pre-SN}}$} & \colhead{$M_{\text{He}}$\tablenotemark{a}} & \colhead{$M_{\text{PNS}}$} & \colhead{$M_{\text{ej}}$}\tablenotemark{b} & \colhead{$\rho_{\text{cental}}$\tablenotemark{c}}  & \colhead{$E_{\text{exp}}$\tablenotemark{c}} & \colhead{$t_{\text{exp}}$\tablenotemark{c}} & \colhead{log$_{10}(L_{\text{peak}})$} & \colhead{$t_{\text{peak}}$\tablenotemark{d}} & \colhead{$t_{-1/2}$} & \colhead{$t_{+1/2}$}}
\startdata
(M$_{\odot}$)    & (cm)             & (M$_{\odot}$)     & (M$_{\odot}$) & (M$_{\odot}$) & (M$_{\odot}$)    & ($g/cm^3$)              & (erg)                & (s)  & (erg $s^{-1}$)            & (day)             & (day)    & (day)\\ \hline
\multicolumn{4}{c|}{Progenitor source: \cite{Sukhbold2016}}                        & \multicolumn{5}{c|}{Explosion source: \cite{Couch2020}}      & \multicolumn{4}{c}{Light curve measurement: This work\tablenotemark{e}}    \\ \hline               
45               & 10.83            & 13.018     & 0.02      & 2.164         & 10.854         & 1.02e+15                & 2.53e+51             & 1.806            & 41.6                      & 99.8             & 53.4  & 77.1         \\
60               & 10.49            & 7.289      & 0.09      & 1.646         & 4.124          & 7.39e+14                & 7.17e+50             & 2.044            & 41.5                      & 77.4             & 43.0  & 90.6         \\
70               & 10.63            & 6.408      & 0.12      & 1.784         & 4.624          & 8.01e+14                & 9.07e+50             & 2.086            & 41.7                      & 66.9             & 37.9  & 63.3         \\
80               & 10.62            & 6.368      & 0.13      & 1.724         & 4.279          & 7.74e+14                & 8.24e+50             & 2.097            & 41.7                      & 75.0             & 44.9  & 72.0         \\
100              & 10.58            & 6.036      & 0.14      & 1.905         & 4.131          & 8.76e+14                & 9.27e+50             & 2.137            & 41.8                      & 51.9             & 27.8  & 59.1         \\
120              & 11.55            & 6.160      & 0.15      & 1.911         & 4.249          & 8.81e+14                & 1.02e+51             & 2.120            & 41.8                      & 51.8             & 27.5  & 58.5         \\ 
\enddata
\tablenotetext{a}{Calculated from \snec composition and density input profile.}
\tablenotetext{b}{Calculated as $M_{\text{ej}} = M_{\text{pre-SN}} - M_{\text{PNS}} - M_{\text{FallBack}}$, where $M_{\text{FallBack}} = 1.518, 0.366$~\Msun for $M_{\text{ZAMS}} = 60, 80$ model, respectively (zero for other models in this table).}
\tablenotetext{c}{Central density, explosion energy, and time since post-bounce at the end of the STIR simulation.}
\tablenotetext{d}{Time since explosion.}
\tablenotetext{e}{The input \Nifs mass is 0.07~\Msun for all models and is mixed out to 60\% of the ejecta in mass coordinate. Note that 0.03 and 0.01~\Msun of \Nifs are trapped at early times in the fallback region for $M_{\text{ZAMS}} = 60, 80$ model, respectively.}

\end{deluxetable*}

\subsection{Light Curve Modeling: SNEC} \label{subsec: Method - SNEC}
The SuperNova Explosion Code \citep[SNEC;][]{Morozova2015}, an open-source 1D radiation-hydrodynamics code\footnote{\url{http://stellarcollapse.org/SNEC}}, is used to further evolve the \stir explosion outcome to homologous expansion stage and simulate the bolometric light curve.
\snec solves Newtonian hydrodynamic evolution assuming spherical symmetry in the Lagrangian frame using the formulation in \cite{Mezzacappa1993}.
Radiation transport in \snec is performed via the equilibrium flux-limited diffusion method and the plasma properties are solved using the stellar equation of state in \cite{Paczynski1983} with a Saha ionization/recombination solver.ing from \stir to \snec are descrisd in \cite{Barker2022}, here we.
Hmma,rize the major adaptations and modifications. 

As the computation domain of \stir only focuses on the explosion itself in the inner region of the star, we combine the structure of the shocked region of the ejecta (\eg, mass, radius, density, velocity) from the final state of \stir simulation with the outer structure of the progenitor star as input for \snec.
The masses of PNS described in Section~\ref{subsec: Method - Explosion} (and listed in Table~\ref{table:pre_post_SN_prop}) are excised as the representation of the remnant in \snec simulations.
The \snec simulation covers from the surface of PNS to the outer edge of the progenitor star in mass coordinate.

The advantage of simulating the explosion physically using \stir instead of artificially using the thermal-bomb or piston-driven explosion in \snec is that the properties of the inner region are driven by physically informed energetics instead of user-input energy over a certain region and time, reducing the systematic uncertainties in the modeling process. 
However, as mentioned in Section~\ref{subsec: Method - Explosion}, the explosion energies of these massive star explosions did not yet reach the asymptotic values when the shock wave approached the $R=\num{15000}$~km radii limit in \stir. 
Hence, we use the analytical approximation of asymptotic explosion energy as \texttt{final\_energy} and propagate the additional energy into the shocked region using the thermal-bomb option with \texttt{bomb\_mode}$=3$ in \snec, see \citet{Barker2022} for details.

The chemical composition is directly mapped from the progenitor star with modification in the shocked region.
Due to the lack of the nuclear reaction network both in \stir and \snec, the true chemical composition of the shocked region is uncertain. 
Following \citet{Barker2022}, we chose to replace the \stir domain with pure $^4$He, which is shown to have neglectable effects on the \snec bolometric light curves (see their Appendix A).
We apply boxcar smoothing on the composition profile four times with a width of 0.4~\Msun each time.
\citet{Morozova2015} has shown that this smoothing procedure, which mimics mixing effects such as Rayleigh-Taylor instability, avoids unphysical bumps in simulated light curves.

\begin{figure*}[htb!]
\centering
\includegraphics[width=0.98\textwidth]{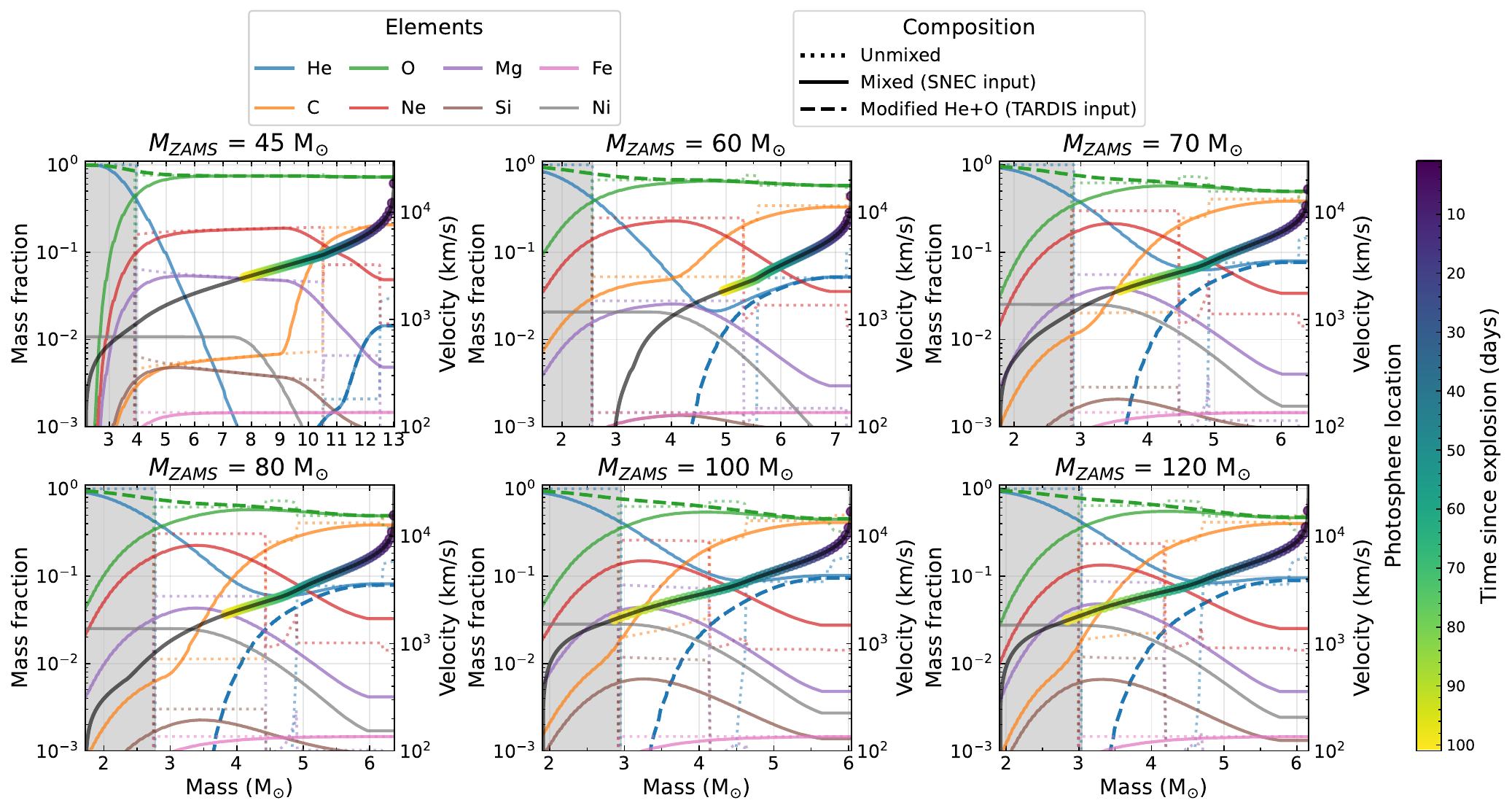}
\caption{Composition and velocity profiles of the models in mass coordinates. The grey region on the left side of each subplot indicates the \stir computation domain, where the \snec input unmixed composition (marked with dotted lines) is replaced with pure He.
The \snec boxcar-smoothed composition is plotted with solid lines.
For TARDIS input composition, the inner pure He is replaced with pure O before the smoothing process, and the modified He and O mass fractions are indicated with dashed lines. 
The velocity profile of each model is presented with a black solid line on the secondary y-axis. 
The photospheric location at a given time is marked with color on the velocity profile.}
\label{fig:snec_composition}
\end{figure*}

We adopt a \Nifs mass of 0.07~\Msun for all our models based on the synthesized value in \citet{Sukhbold2016} using parametrized explosion models, as neither \stir nor \snec has nuclear-burning networks \citep[following][]{Barker2022}.
\snec offers customized mixing degrees for \Nifs, and we choose to mix it out to 60\% of \snec domain in mass coordinates.

We treat the material that has zero expanding velocity in the inner region at 1~d past explosion in \snec as potential fallback mass $M_{\text{FallBack}}$.
Then the ejecta mass of the model ($M_{ej}$) is calculated through $M_{\text{ej}} = M_{\text{pre-SN}} - M_{\text{PNS}} - M_{\text{FallBack}}$.
Half the models are fully unbound with zero $M_{\text{FallBack}}$.
The two models with $M_{\text{ZAMS}} = 50, 55$~\Msun have significant $M_{\text{FallBack}}$ ranging from 4 to 5~\Msun, which may form black holes after the weak explosion. 
The other two models with $M_{\text{ZAMS}} = 60, 80$~\Msun have $M_{\text{FallBack}} = 1.5, 0.4$~\Msun, respectively.
Table~\ref{table:pre_post_SN_prop} tabulates the ejecta mass of selected models.

The \snec bolometric luminosity output is composed of the radiative luminosity at the photosphere and the integrated luminosity from $\gamma$-ray (produced by radioactive \Nifs) deposition above the photosphere \citep[see][]{Morozova2015}. 
\snec determines the photosphere position by identifying the location when optical depth $\tau = 2/3$, which is tracked by acquiring the composition- and density-dependent Rosseland mean opacities.
Fig.~\ref{fig:snec_composition} shows the photospheric location and composition of the selected models as a function of mass coordinates.

\subsection{Spectral Modeling: TARDIS} \label{subsec: Method - TARDIS}

We pass the resultant bolometric luminosity, photosphere location, and ejecta profiles (i.e., velocity, density, and composition) from \snec to the open-source 1D Monte Carlo radiative transfer code \tardis\footnote{\url{https://github.com/tardis-sn/tardis}} (\citealp{Kerzendorf2014}; version: \citealt{kerzendorf_2024_13207705}), to simulate the spectral time series. 
\tardis self-consistently solves the plasma properties (e.g., radiative temperature, dilution factor, ionization/excitation state, and electron density) at a given time snapshot.
This is done by iteratively propagating indivisible energy packets that follow a blackbody profile from the photosphere (the inner boundary) through a spherically symmetric and homologously expanding ejecta till the plasma state is converged. 
We limit our spectral simulations to 75 to 115 days after the explosion when less than 20\% of the \snec luminosity is contributed by the energy deposited by $\gamma$-ray above the photosphere.

The key configuration settings used for \tardis simulation are tabulated in Table~\ref{table:tardis_settings}, of which the detailed description can be found in \citet{Kerzendorf2014}.
Helium is treated with ``\texttt{recomb-NLTE}" mode \citealt{Boyle2017}, which is an analytic approximation for the full statistical equilibrium calculations with non-thermal effects described in \citealt{Hachinger2012}.
\citet{Hachinger2012} found that the vast majority of the helium population remains singly ionized in their SNe~Ib/c models, and the \ion{He}{1} excited states are strongly coupled with \ion{He}{2} ground state.
The ``\texttt{recomb-NLTE}" mode utilizes such coupling and calculates the helium ions/level population relative to the population of \ion{He}{2} ground state, assuming negligible \ion{He}{1} ground states.
Such approximation is shown to be able to reproduce similar spectral results of both modeled and observed SESNe \citep[e.g.,][]{Williamson2021, Kwok2022, Agudo2023}.

We adopt the bolometric luminosity and photospheric velocity from \snec output as the requested luminosity and inner boundary velocity in \tardis at a given time, respectively.
\tardis requires density and composition profiles as a function of velocity.
We first construct a velocity grid with 45 shells within the effective velocity range of the \snec output.
Then, we interpolated the density and boxcar smoothed composition profile processed by \snec (see Section~\ref{subsec: Method - SNEC}) from mass coordinates to this velocity grid using the mass-velocity output profile at each time step.
In addition, the radiative temperatures from the \snec output are also mapped into the configuration as the initial guess for the Marto Carlo process in \tardis.

While replacing the composition within the \stir domain with pure helium does not impact the bolometric light curve produced by \snec, it will affect the synthetic spectra generated by \tardis if the photosphere reaches the inner region.
Hence, we replaced the pure He with O before the boxcar smoothing procedure for composition mapping modification only (other properties are kept the same from the original \snec output. 
This only affects the composition of He and O in the inner regions, which does not affect the early time spectra when the photosphere is above the modified location.

\begin{deluxetable}{l|l}[t!]
\centering
\tablecaption{Key configuration settings in \tardis simulations \label{table:tardis_settings}}
\tablehead{
\colhead{Configuration name} & \colhead{Setting}}
\startdata
Atomic data          & kurucz\_cd23\_chianti\_H\_He.h5 \\
Ionization           & nebular                     \\
Excitation           & dilute-lte                  \\
Radiative rate       & dilute-blackbody            \\
Line interaction     & downbranch                  \\
Helium treatment     & recomb-nlte                 \\
Number of iterations & 30                          \\
Number of packets\tablenotemark{a}    & $10^5$     \\ 
\enddata
\tablenotetext{a}{Number of packets used for the last iteration is increased to $10^6$.}
\end{deluxetable}

\section{Results and Discussion} \label{sec: results}

We present synthetic light curves and spectra of physics-driven explosions of six H-free/He-poor WR progenitor stars that exploded in C20.
The model light curves and spectra, along with the \snec and \tardis configuration files, are publicly available on Zenodo\footnote{\href{https://zenodo.org/records/14343229}{doi: 10.5281/zenodo.14343229}}.
We exclude the two models with $M_{\text{ZAMS}} = 50$ and $55$~\Msun from the analysis because they contain a significant amount of bound material in the inner ejecta ($4-5$\Msun), which could artificially increase the late-time luminosity \citep[\eg,][]{Dexter2013}.

\subsection{Caveats in Progenitor Adoption}
The progenitor property is a high-dimension problem that is sensitive to not only the main sequence mass, but also metallicity, rotation, and binary effects.
The choice of physics assumptions, such as mass loss treatment and convection prescription, affect not only the outer layer but also the core structure of the progenitor star \citep[\eg,][]{Woosley1993, Dewi2002, Farmer2016, Renzo2017, Ertl2020, Laplace2021, Schneider2021}.
We note that the mass loss rate adopted in \citet{Sukhbold2016} is stronger than the common prescription in recent studies \citep[\eg,][]{Nugis2000, Smith2014, Antoniadis2024}. 
Furthermore, numerical resolution in mass separation and nuclear burning networks can cause significant variations in stellar properties \citep[\eg,][]{Farmer2016, Renzo2024}.

To self-consistently explore how progenitor system variations affect SN observables, it is essential to understand how limitations in each stage of the sequential simulation framework influence the results.
The absence of nuclear-burning networks in \stir and \snec hinders a direct connection between progenitor properties and the \Nifs yield, which is currently fixed at 0.07~\Msun (see Section~\ref{sec: methods}). 
The mixing of \Nifs, which strongly affects the light curve shape and helium spectral features (see Sections~\ref{subsec: results - Bolo_LC} and \ref{subsec: results- He features}), is also fixed due to the limitation of the 1D framework. 
Additionally, the difference in opacity prescriptions between \snec and \tardis is another source of uncertainties, such as the determination of photospheric location can be inconsistent.
Furthermore, the current helium treatment in \tardis approximates the detailed, balanced level population solution under the assumption that most of He is singly ionized (see Section~\ref{subsec: results- He features}). 
Finally, a detailed comparison of how different codes influence outcomes at each simulation stage -- stellar evolution, explosion, and radiative transfer --  is highly encouraged, though it is beyond the scope of this study.

\subsection{Bolometric Light Curve} \label{subsec: results - Bolo_LC}

\begin{figure}[!htb]
\centering
\includegraphics[width=0.93\columnwidth]{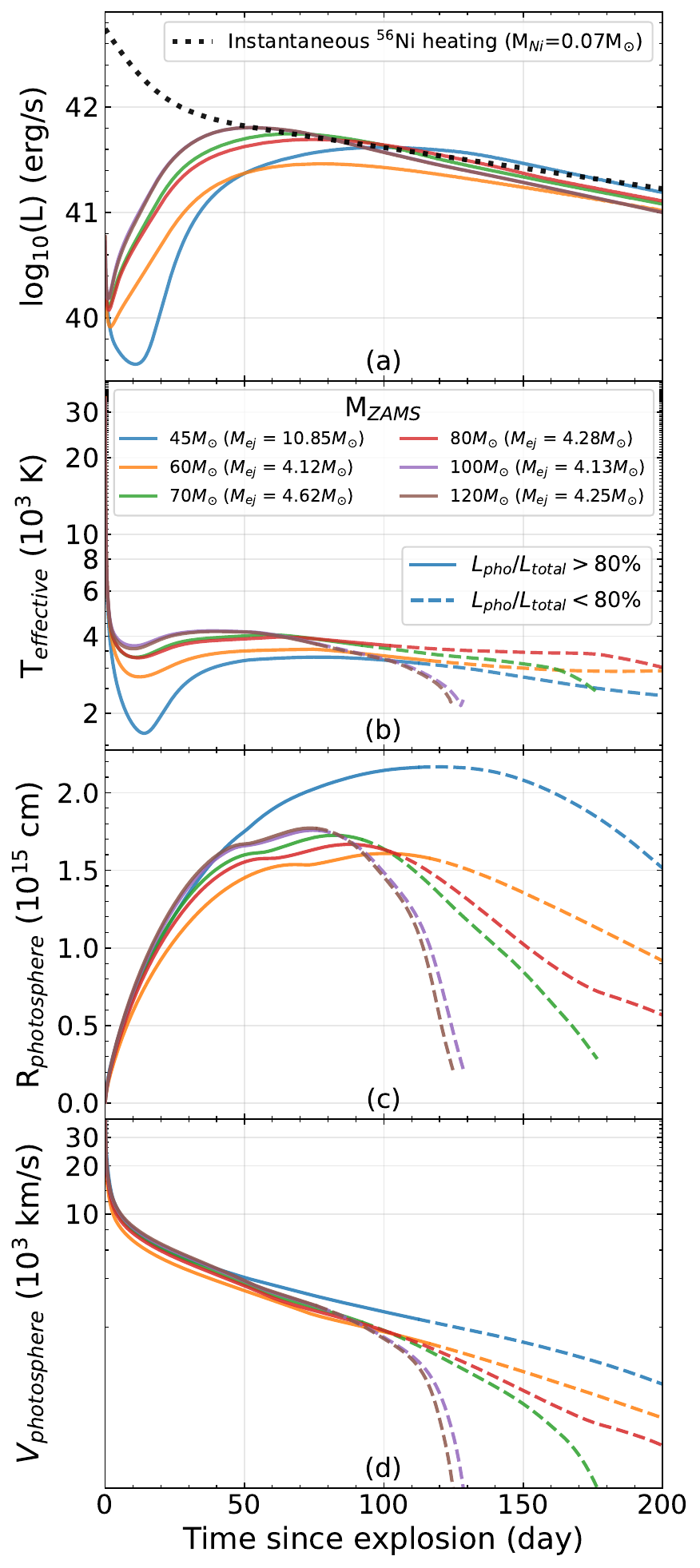}
\caption{The \snec bolometric light curves (a) of the models and their photospheric effective blackbody temperature (b), photospheric radius (c), and photospheric velocity (d). The lines are color-coded by the $M_{\text{ZAMS}}$ of the progenitors. The dotted line in panel (a) represents the analytical form of the instantaneous radioactive heating of \Nifs from \citet{Wheeler2015} assuming the $\gamma$ diffusion time scale is infinity. In the lower three panels, the photospheric phase ($L_{\text{photosphere}}/L_{\text{total}} \ge 80\%$) is indicated with solid lines. We do not compute \tardis spectra beyond the photospheric phase.}
\label{fig:SNEC_LC_and_photosphere_info}
\end{figure}

Figure~\ref{fig:SNEC_LC_and_photosphere_info} presents the \snec bolometric light curves, along with the time evolution of photospheric parameters, including effective temperature, radius, and expansion velocity. 
Table~\ref{table:pre_post_SN_prop} tabulates the measured bolometric light curve properties. 

\begin{figure}[!htb]
\centering
\includegraphics[width=0.99\columnwidth]{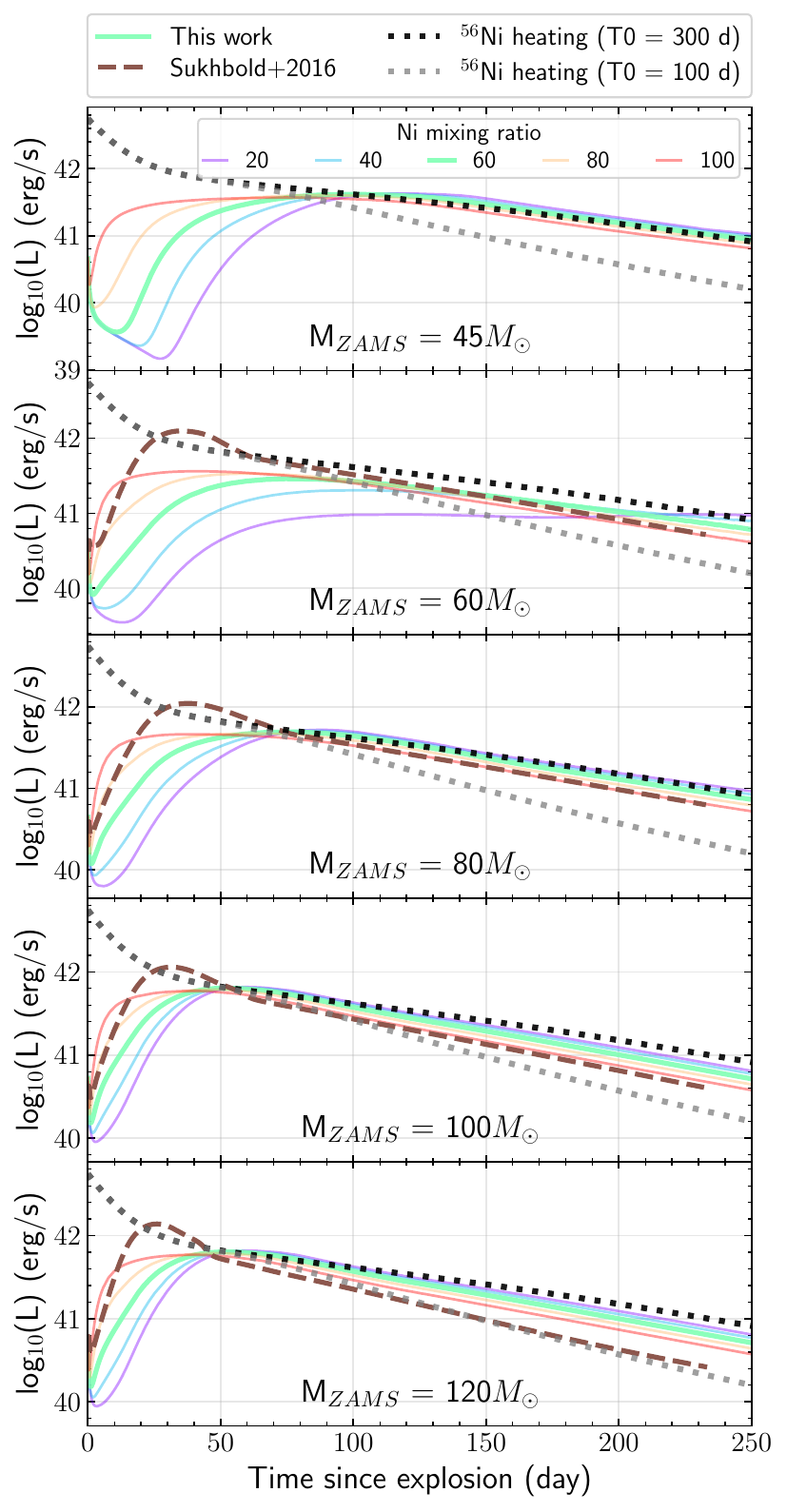}
\caption{The \snec bolometric light curves of models with varying \Nifs mixing percentage (relative to the total ejecta mass in mass coordinate).
The light curves of the same set of progenitors exploded in \citet{Sukhbold2016} are plotted for comparison, except the $M_{ZAMS} = 45$~\Msun model, which did not explode in their work.
The dotted black and grey lines are thermalized heating energy rates from radioactive decay of \Nifs = $0.07$~\Msun following the analytical models in \citet{Wheeler2015}, with a $\gamma$-ray diffusion time scale of 300 and 100 days, respectively.} 
\label{fig: SNEC_LC_compare_S16}
\end{figure}

The bolometric light curves of our models have a $t_{-1/2}$ ranging from 30 to 50 days and a $t_{+1/2}$ ranging from 50 to 90 days, where $t_{-1/2}(t_{+1/2})$ indicates the time scale between the peak and the half peak luminosity before (after) the maximum. 
The ones with larger ejecta masses tend to have broader bolometric light curves, consistent with the analytical diffusion model prediction \citep{Arnett1982} and simulated SESN light curves from \citet{Woosley2021}.
The model with the largest ejecta mass ( $M_{\text{ZAMS}} = 45$~\Msun) has the longest rise time ($\sim$100~days).
We note that the rise time is sensitive to the mixing degree of \Nifs, with stronger mixing leading to a shorter rise time, see Fig.~\ref{fig: SNEC_LC_compare_S16}.
The rise time of our models can be reduced by a few tens of days if \Nifs is fully mixed, while the peak luminosity stays consistent with only a few percent modification.
We compared our model light curves to observations and found that they are much broader than those of the typical SESNe population, which has median $t_{-1/2}(t_{+1/2})$ value on the order of 10 (20)~days \citep[\eg,][]{Prentice2019}.
The rise/decline rates of our models marginally resemble those of observed SESNe with broad light curves, which have $t_{-1/2}(t_{+1/2})$ values of $\sim$ 25 (50)~days and may account for more than 10\% of the SESN population \citep[see \eg][]{Karamehmetoglu2023}. 

The peak bolometric luminosity ($L_{\text{peak}}$) of our models covers a range of 10$^{41.5} - 10^{41.8}$~\Lunit, which is on the lower end compared to the average values inferred from observations \citep[10$^{42} - 10^{43}$~\Lunit;][]{Lyman2016, Taddia2018}. 
The modeled light curves of these high-initial-mass progenitors stars being broad and faint is consistent with previous modeling efforts \citep[\eg,][]{Ensman1988, Ertl2020, Woosley2021}. 
Low-mass progenitors that are stripped through binary effects are believed to be the primary source of normal SESNe \citep[\eg,][]{Dessart2011, Smith2014, Yoon2015}.
However, the underproduction of \Nifs in numerical simulations compared to the values inferred from observations persists regardless of the progenitors \citep[\eg,][]{Anderson2019, Sawada2023}, which may suggest the existence of additional power sources in SESNe \citep[\eg,][]{Afsariardchi2021, Rodriguez2023, Rodriguez2024}.

A noticeable difference in the bolometric light curve shape among the models is that the initial post-breakout peak in the $M_{\text{ZAMS}}$ = 45~\Msun model is accompanied by a long declining phase ($\sim$15~d) before rising to the main luminosity peak driven by the radioactive \Nifs, while it only lasts up to a few days in other models. 
This initial decline in luminosity is related to the rapid cooling of the photosphere during early time, as shown in panel (b) of Fig.~\ref{fig:SNEC_LC_and_photosphere_info}, which presents the effective blackbody temperature of the photosphere calculated using Stefan-Boltzmann law. 
The luminosity and duration of the shock cooling emission phase are sensitive to the ejecta structure, composition, and mixing effects \citep[\eg,][]{Woosley1994, Dessart2011, Bersten2012, Nakar2014, Piro2021, Curtis2021}. 
Stronger \Nifs mixing leads to more heating in the outer layers from the \Nifs decay to compensate for cooling effects due to ejecta expansion, which results in a shorter shock cooling phase, see Fig.~\ref{fig: SNEC_LC_compare_S16}.

At 100 $-$ 200~days after the explosion, the luminosity is dominated by the heating deposited by the $\gamma$-ray from the \Nifs decay products. 
The declining tail of the bolometric light curve shows noticeable diversity despite all models having the same \Nifs mass.
This difference is primarily caused by the variation of the $\gamma$-ray diffusion time scale that controls how efficiently the $\gamma$-ray can contribute to the heating of the ejecta \citep[][and references within]{Wheeler2015}.
The models with smaller ejecta mass have steeper decline slopes at late times, corresponding to less effective $\gamma$-ray trapping by the ejecta.
Comparison of true and inferred ejecta mass from bolometric light curves is discussed in Section~\ref{subsec: results - M_ej from LC}.

Figure~\ref{fig: SNEC_LC_compare_S16} also compares the bolometric light curve with the ones calculated in \citet{Sukhbold2016} if available, which were simulated using \kepler based on the explosion outcomes calibrated to the central engine N20 model \citep{Shigeyama1990, Nomoto1988, Saio1988}.
The discrepancy arises from the difference in the numerical methods and parameters.
The rising part of the light curve is sensitive to the opacity floor setting that is commonly adopted in numerical simulations that can account for additional opacity sources \citep{Haynie2023} (hereafter H23).
The opacity floor value adopted in \citet{Sukhbold2016} is set to be 10$^{-5}$~cm$^2$~g$^{-1}$ in \kepler.
In \snec simulation, the fiducial opacity floor setting consists of 0.24 and 0.01~cm$^2$~g$^{-1}$ in the metal-rich core and metal-poor envelope region, respectively, which returns a metallicity-dependent opacity floor value in each simulation grid. 
A lower opacity floor value allows more recombination, which decreases the effective mass available for diffusion in the outer ejecta. 
This causes the rising part of the light curves to evolve faster and reach a brighter peak (see H23).
The declining tails depend on both ejecta and \Nifs mixing, which also differs between our models and \citet{Sukhbold2016}.

\subsection{Ejecta Mass Comparison} \label{subsec: results - M_ej from LC}

The model ejecta mass ($M_{ej}$) is calculated through $M_{\text{ej}} = M_{\text{pre-SN}} - M_{\text{PNS}} - M_{\text{FallBack}}$ and listed in Table~\ref{table:pre_post_SN_prop}, where $M_{\text{FallBack}}$ is the mass fallback onto the renmant approxiamated from \snec output. 
The material with zero expanding velocity in the inner region at 1~day past explosion is treated as $M_{\text{FallBack}}$.
Most of the selected models are fully unbound except for two models with $M_{\text{ZAMS}} = 60, 80$~\Msun, which have $M_{\text{FallBack}} = 1.52,\,0.37$~\Msun, respectively.
The resulting ejecta masses range from 4 to 11~\Msun and are tabulated in Table~\ref{table:pre_post_SN_prop}.
This range is on the higher end compared to the estimated distribution from normal SESNe observations \citep[\eg,][]{Lyman2016, Prentice2019, Rodriguez2023} but consistent with the SESNe that have broad light curves \citep[\eg,][]{Taddia2019, Karamehmetoglu2023}. 

Here, we compare the model ejecta mass with the inferred values from the synthetic observations following the analytical relations compiled in \citet{Wheeler2015} and H23.
\citet{Wheeler2015} compiled two commonly used methods that infer from the rising and declining part of the light curve following \citet{Arnett1982} and \citet{Clocchiatti1997}, respectively.
H23 calibrated the light curve tail formulation from \citet{Wheeler2015} based on a grid of parametrized explosion models of a set of intermediate-mass binary progenitors published in \citet{Laplace2021} using \snec. 
Table~\ref{table: M_ej methods} summarizes the formalisms mentioned above for measuring ejecta mass, where $\kappa$ and $\kappa_{\gamma}$ represents the effective optical opacity and $\gamma$-ray opacity, with a fiducial value of 0.1 and 0.03~$cm^2\,g^{-1}$, respectively.

The rise time ($t_r$) is measured to be the time from the explosion to the peak of the bolometric light curves, same as the $t_{\text{peak}}$ value listed in Table~\ref{table:pre_post_SN_prop}.
The photospheric velocity ($v_{\text{ph}}$) at the peak luminosity is directly taken from the \snec simulation output. 
To measure the $\gamma$-ray diffusion time ($T_0$), we follow Eq.~(2) and (5) from H23 to fit the late-time light curve tail to the thermalized energy deposited rate given by:
\begin{equation}
L_{\text {heat }}=L_{\mathrm{Ni}}\left[1-e^{-\left(T_0 / t\right)^2}\right] \text {, }
\end{equation}
where $t$ is time since explosion and $L_{\mathrm{Ni}}$ represents the instantaneous radioactive heating rate from \Nifs decay: 
\begin{equation}
L_{\mathrm{Ni}}(t)=\frac{M_{\mathrm{Ni}}}{M_{\odot}}\left[\epsilon_{\mathrm{Ni}} e^{-t / \tau_{\mathrm{Ni}}}+\epsilon_{\mathrm{Co}}\left(e^{-t / \tau_{\mathrm{Co}}}-e^{-t / \tau_{\mathrm{Ni}}}\right)\right] {. }
\end{equation}
Here $\epsilon$ and $\tau$ are the decay energy and decay time of the corresponding isotope, respectively. 
The fitting process is performed using the \texttt{curve\_fit} function in the Python package \texttt{Scipy} \citep{2020SciPy-NMeth}.

\begin{figure}[!htb]
\centering
\includegraphics[width=\columnwidth]{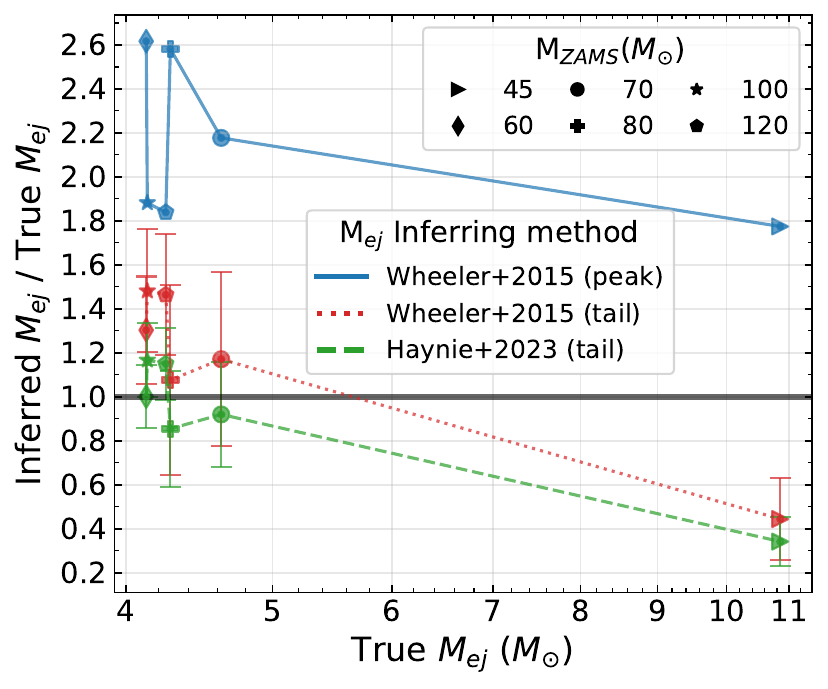}
\caption{The ratio of the inferred ejecta mass from bolometric light curves and the model ejecta mass from simulation.
The inferred ejecta masses are based on the analytical models summarized in \citet{Wheeler2015} and \citet{Haynie2023}.
The error bars on the ejecta mass ratios that are calculated based on the light curve tail represent the variation in measured $\gamma$-ray diffusion time scale $T_0$, resulting from different fitting time ranges.  }
\label{fig:LC_M_ej_inferred}
\end{figure}

The ratio of the inferred over the true model ejecta mass is plotted in Fig.~\ref{fig:LC_M_ej_inferred}. 
Note the ``errorbar" on the two light-curve tail methods indicates the range of the $M_{ej}$ ratio using $T_0$ values fitted from difference time ranges (a moving 50~$d$ window from 150 to 300 days).
The markers represent the average value of the calculated $M_{ej}$ ratio.
The light-curve peak method systematically overestimates the ejecta mass by 80\% to 160\%, while the two light-curve tail methods have a maximum uncertainty of 80\%.

One potential cause of the difference is the assumption of a constant effective opacity in the analytical model, which in nature varies in both space and time \citep[\eg,][]{Dessart2015, Wheeler2015, Dessart2016}.
Especially in SESNe ejecta, the recombination front of H and/or He in the outer layer quickly decreases opacity \citep{Dessart2016}.
Furthermore, it is not clear if the effective opacity is the same among different SESN progenitors \citep[\eg,][]{Lyman2016, Dessart2016, Haynie2023} as opacity is sensitive to several factors, including temperature, density, composition, and dynamical properties of the ejecta \citep{Hoeflich1993, Pinto2000}.
\citet{Wheeler2015} suggested that the commonly assumed mean opacity around the peak is overestimated. 

Overestimating the rise time from the model light curves is another potential reason for the large discrepancy of the ejecta mass inferred from the rising light curves.
The rising part of the bolometric light curves is susceptible to the \Nifs mixing and opacity floor setting in models has H and/or He since the floor values effectively reflect the recombination rates \citep{Dessart2016, Haynie2023}, see Section~\ref{subsec: results - Bolo_LC}.
The systematic overestimation using the rising light curves may suggest that the fiducial opacity floor setting in \snec, initially calibrated for SNe~II \citep{Bersten2011}, may be too high for SESNe.

The light-curve tail method from H23 with calibrated parameters performs best, particularly in cases where the true ejecta mass is below 5~\Msun.
This range aligns with the calibration sample used in H23, which consisted of intermediate-initial-mass progenitors with ejecta masses between 2 and 5~\Msun.
For our models that fall within this ejecta mass range, the average discrepancy remains within 20\%.
However, for models with higher ejecta masses, such as the one with $M_{\text{ej}} = 10.85$~\Msun (corresponding to the progenitor with $M_{\text{ZAMS}} = 45$~\Msun), the H23 method underestimates the ejecta mass by about 70\%.
The inferred mass for this model is only 4.9~\Msun, which falls within the typical ejecta mass range of SESNe inferred from observed samples \citep[\eg,][]{Lyman2016, Prentice2019, Rodriguez2023}.
This discrepancy raises concerns about the potential bias in estimating higher ejecta-mass models, as the calibrated method might systematically underestimate the true ejecta mass.

While our sample size is too limited to draw definitive conclusions, this result suggests that caution should be exercised when applying analytical models to high ejecta-mass SESNe. 
Further exploration with a more extensive grid of progenitor models and better characterization of the effects of \Nifs mixing and opacity variations would be valuable for refining the accuracy of ejecta mass estimation from SESN light curves. 
While analytical methods provide a useful first-order estimate, more sophisticated modeling is needed to fully capture the complexity of these explosions and their observable properties.

\begin{deluxetable*}{l|l|l|l}[!htb]
\centering
\tablecaption{Summary of methods used to measure SN ejecta mass from bolometric light curve} \label{table: M_ej methods}
\tablehead{ & Light curve peak & Light curve tail & Light curve tail}
\startdata
Reference           & \citet{Wheeler2015}    & \citet{Wheeler2015}  & \citet{Haynie2023}        \\ \hline
\multirow{2}{*}{Analytical Equation\tablenotemark{a}}
& \multirow{2}{*}{$M_{ej}$ = $\frac{1}{2} \frac{\beta c}{\kappa} v_{\text{ph}} t_{r}^2$}
& \multirow{2}{*}{$M_{ej}$ = $\frac{3}{10} \frac{v_{\text{ph}}^2 T_0^2}{\eta \kappa_{\gamma}}$}
& \multirow{2}{*}{$M_{ej}$ = $10^{-\frac{A}{B}} v_{\text{ph}}^{-\frac{C}{B}} T_0^{\frac{1}{B}}$}     \\ 
 &  &  &  \\ \hline
Constants           
& \begin{tabular}[c]{@{}l@{}}$\beta$ = 13.8\\ $\kappa = 0.1 \,cm^2\,g^{-1}$ \\ c = 3$\times10^8 \,cm\,s^{-1}$ \end{tabular} 
& \begin{tabular}[c]{@{}l@{}}$\eta$ = 0.05 \\ $\kappa_{\gamma} = 0.03 \,cm^2\,g^{-1}$\end{tabular} 
& \begin{tabular}[c]{@{}l@{}}A = $-$4.34 $\pm$ 0.65\\ B = 0.608 $\pm$ 0.02 \\ C = $-$1.05 $\pm$ 0.02\end{tabular} \\ \hline
\enddata
\tablenotetext{a}{$v_{\text{ph}}$ indicates the photospheric velocity, $t_r$ indicates the rise time to the bolometric luminosity peak, and $T_0$ indicates the $\gamma$-ray diffusion time measured from the bolometric light curve declining tail.}
\end{deluxetable*}

\subsection{Spectral Time Series} \label{subsec: results - spectra}

\begin{figure*}[t!]
\centering
\includegraphics[width=\textwidth]{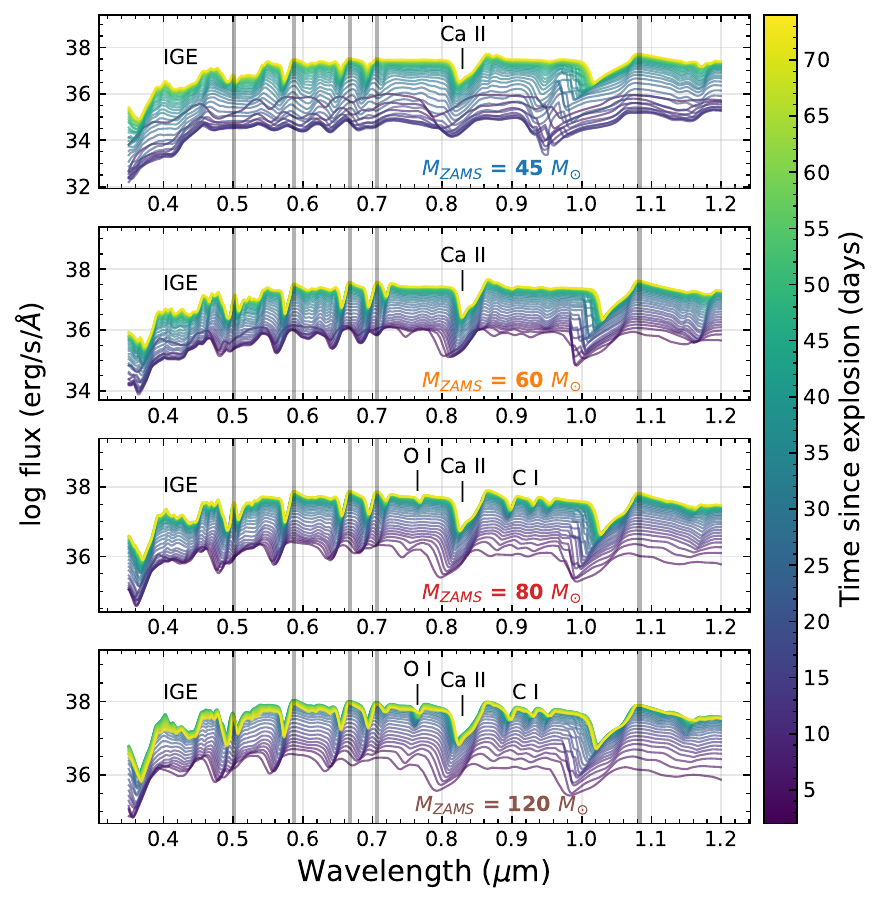}
\caption{\tardis spectral time series of $M_{\text{ZAMS}}$ = 45, 60, 80, 120 \Msun model. Spectra are color-coded with time relative to the explosion. The vertical dark gray lines mark the strong He lines in the restframe.} 
\label{fig:tardis_spec_time_series}
\end{figure*}

\begin{figure*}[htb!]
\centering
\includegraphics[width=0.95\textwidth]{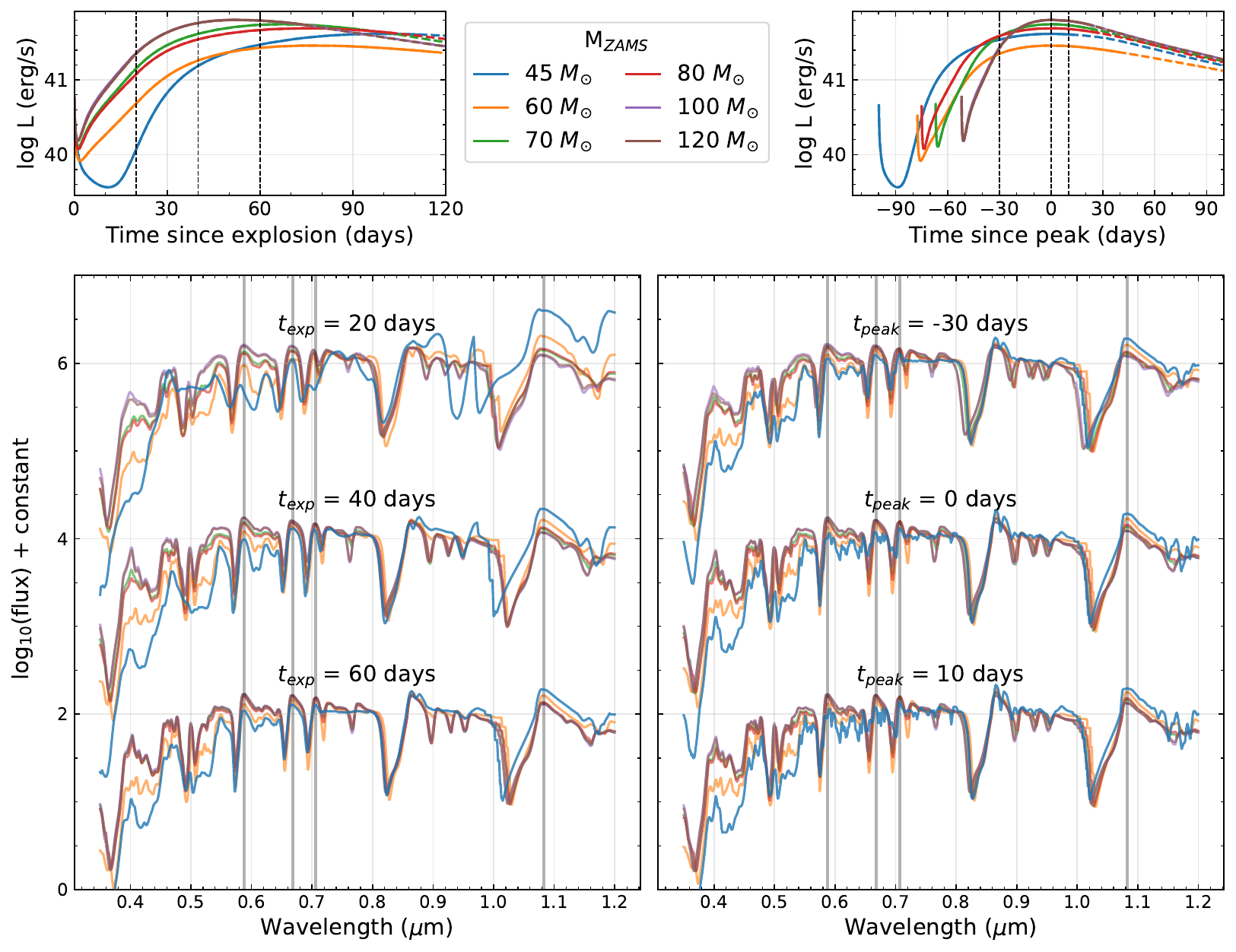}
\caption{\tardis spectral comparison at fixed phases relative to the explosion date (left column) or the bolometric light curve peak (right column). On the top row, the bolometric light curves are plotted for reference purposes with the corresponding time reference as the column. The dashed vertical lines in the top panels mark the selected phases for spectral comparison. The flux is normalized by dividing the mean flux within the wavelength range from 7000 to 9000 $\AA$ and offset by a constant. The vertical dark gray lines mark strong He lines in the restframe.}
\label{fig:tardis_spec_comp_at_fixed_times}
\end{figure*}

\begin{figure*}[bt!]
\centering  
\includegraphics[width=0.98\textwidth]{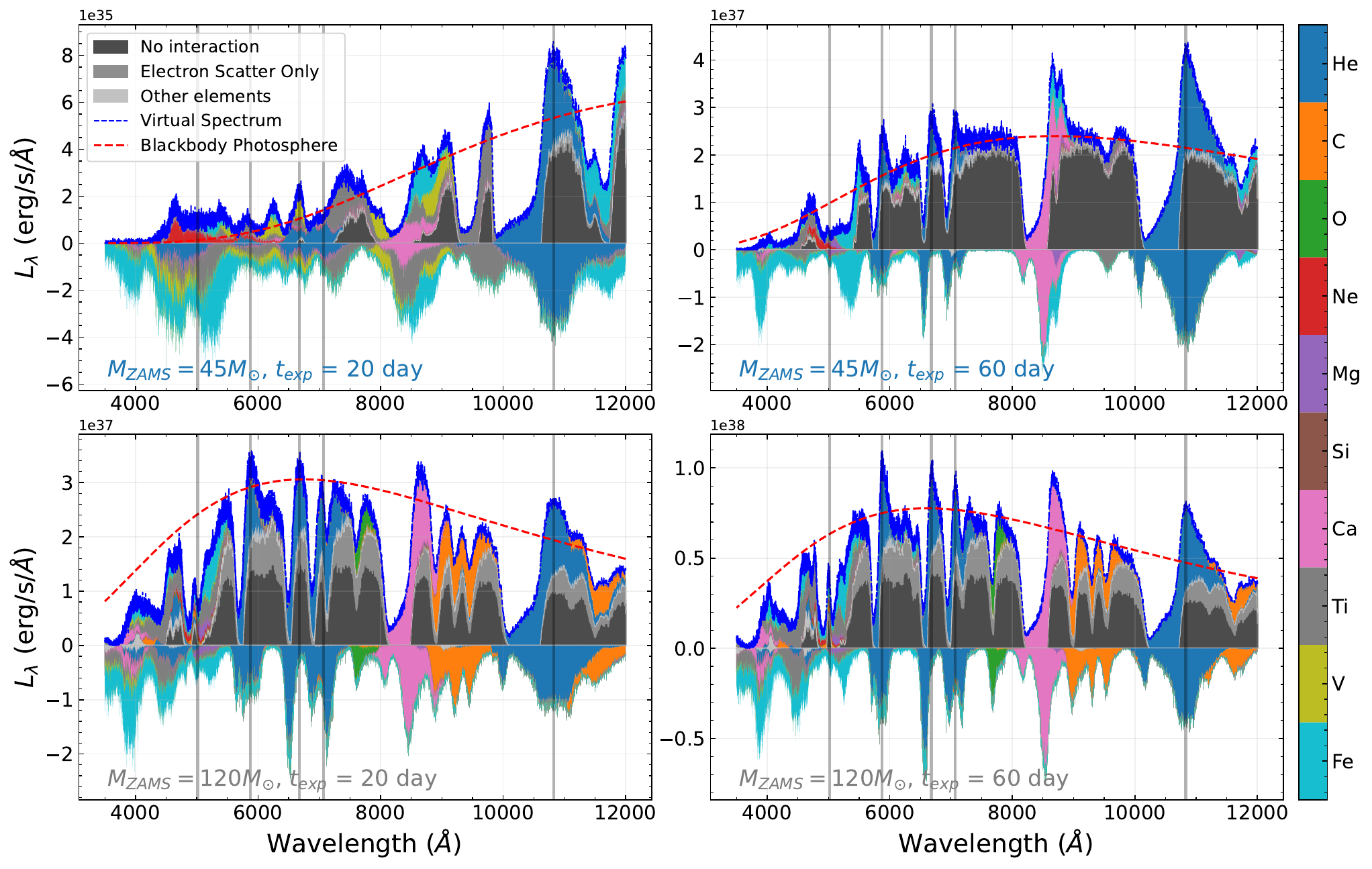}
\caption{\tardis elemental deposition plot of $M_{\text{ZAMS}}$ = 45~\Msun (top panels) and 120~\Msun (bottom panels) model at 20 (left panels) and 60 (right panels) days. The colored patches are the cumulated wavelength distribution of energy packets before (indicated with negative flux) and after (indicated with positive flux) the last interaction in the \tardis simulation. The simulated spectrum is the resultant collection of the energy packets that escaped from the ejecta after the last interaction, marked with a dashed blue line. The vertical dark gray lines mark strong He lines in the restframe.}
\label{fig:TARDIS_SDEC_plot}
\end{figure*}

Spectra provide important clues to the chemical composition and dynamical structure of the SN ejecta, which can be directly linked to the progenitor and explosion itself. 
However, the detailed and publically available radiative transfer spectral models of stripped massive star explosions is limited, especially for stars on the higher mass end.
The model spectral time series presented in this work expands the quantities and diversity of the currently available pool of model spectra of plausible SESNe progenitors.

The synthetic spectral time series of our models during the photospheric phase from \tardis simulations are showcased in Fig.~\ref{fig:tardis_spec_time_series}.
In all models, most absorption features evolve to be narrower over time.
He lines are visible in all models throughout the photospheric phase, with a more detailed discussion presented in Section~\ref{subsec: results- He features}.
Regardless of the presence of He lines, we do not explicitly label the models as SNe~Ib or SNe~Ic candidates, since the line formation of He is not only dependent on the ejecta mass profile but also the radiation field that is sensitive to the distribution of \Nifs \citep[\eg,][]{Dessart2011}, which is a fixed input in this work.

When comparing the model spectra at a fixed time relative to the explosion or peak in Fig.~\ref{fig:tardis_spec_comp_at_fixed_times}, it is shown that the low ejecta mass group ($M_{\text{ZAMS}}$ = 60, 70, 80, 100, and 120 \Msun) are consistently similar to each other in spectral features regardless of the choice of reference time frame. 
The high eject mass model ($M_{\text{ZAMS}}$ = 45 \Msun) shows distinct features at early times in the frame relative to the explosion, while appearing spectroscopically similar to the low eject mass group in time relative to the peak.
This is largely due to the temperature difference as the $M_{\text{ZAMS}}$ = 45 \Msun model has a significantly longer cooling phase and the ejecta temperature is consistently lower during the early time, see Section~\ref{subsec: results - Bolo_LC}.
Early time spectral observations of SESNe are valuable to differentiate the ejecta properties \citep[\eg,][]{Williamson2023, Yesmin2024}.

Figure~\ref{fig:TARDIS_SDEC_plot} presents the Spectral element DEComposition (SDEC) plot from \tardis simulation, demonstrating the spectral features contributions based on the last interaction of the energy packets within the simulation.
The $M_{\text{ZAMS}}$ = 45 \Msun model at 20~days past explosion (upper left panel) shows weak He features blended with other ions. 
At this time, the spectral line profiles are largely dominated by \ion{Ti}{1}, \ion{V}{1}, and \ion{Fe}{1} in the optical region, including the broad absorption feature around 8000~\AA, which is solely dominated by Ca in the lighter ejecta mass model ($M_{\text{ZAMS}}$ = 120 \Msun) shown in the bottom left panel. 
Most of the elements are not ionized due to the low radiative temperature at this time. 
No C or O lines are present in the spectrum, despite them being the most abundant elements in terms of mass fraction in this model, likely due to the relatively low temperature.
Note that at low temperatures, the effect of the wavelength-dependent photosphere is larger and poses additional uncertainties on the continuum flux of simulated TARDIS spectra (O'Brien et al. in prep.).

The $M_{\text{ZAMS}}$ = 120 \Msun model, on the other hand, shows line profiles that are consistent of \ion{He}{1}, \ion{C}{1}, and \ion{Ca}{2} at the same epoch relative to the explosion (see the bottom left panel in Fig.~\ref{fig:TARDIS_SDEC_plot}). 
The line profiles remain consistent in ion contribution but become slower in velocity and narrower in broadness when approaching the bolometric light curve maximum (see the lower right panel).
While the $M_{\text{ZAMS}}$ = 45 \Msun model shows a more significant change in ion contribution over time.
The difference is primarily driven by the variation within the radiation field and the resultant plasma state.

\citet{Dessart2015} has suggested that the strong presence of \ion{C}{1} lines near 1~$\mu$m region could indicate underabundant He in the progenitor. 
This is consistent in the low-ejecta-mass models in our work, with a maximum He mass fraction of $\sim$10\% in the outer layer.
However, the absence of \ion{C}{1} lines along with the strong presence of \ion{He}{1} lines does not necessarily indicate the presence of a pure He shell as mentioned in \citet{Dessart2015}.
Here, we caution that the absence of \ion{C}{1} feature could also arise from low radiative temperature instead of abundance effect. 
We note here that the radiation field and plasma state are dependent on the \Nifs input, the \ion{C}{1} feature can be strongly present in the same model with increased \Nifs mass, which increases the radiation temperature.

\subsection{Helium Features} \label{subsec: results- He features}
The strength of the He feature is not directly correlated with He mass in the ejecta, but rather strongly dependent on a combination of factors, including the radiation field and composition of the ejecta \citep[\eg,][]{Dessart2012, Hachinger2012}.
The He atom has high ionization and excitation energy that requires non-thermal effects (such as high-speed electrons Compton scattered by $\gamma$-ray) for line production if the radiation temperature is not sufficiently high \citep[\eg,][]{Harkness1987, Lucy1991, Kozma1992, Dessart2011, Hachinger2012}.
The non-thermal effects come from radiative decay products of \Nifs situated near the line-forming region. 
The mass fraction of He also can affect the line strength. 
Assuming a fixed amount of non-thermal electrons at line forming region, a mixed He layer can produce weaker He features compared to a pure He layer since there is less energy contributed to He proportionally \citep[\eg,][]{Dessart2012, Dessart2015}.

Several spectral studies have attempted to determine the hidden helium mass in models representing SNe Ib/c, but no consensus has been reached yet. 
\citet{Hachinger2012} determined an upper limit of 0.14~\Msun of He based on ejecta models with similar stellar cores but different He envelopes.        
\citet{Dessart2015} showed that more He can be hidden (0.3~\Msun) in models with more C/O mixed outward, using piston-driven explosions of binary progenitors from \citet{Yoon2010}.
\citet{Teffs2020} demonstrated that the He line detection limit is lower in near-infrared (NIR) compared to optical, and 0.02~\Msun or less He can saturate the \ion{He}{1} 1.083~$\mu$m line, based on parametrized explosions of a $M_{\text{ZAMS}}$ = 22~\Msun progenitor model.
\citet{Williamson2021} found an upper limit of 0.05~\Msun based on optical \ion{He}{1} lines for a case study of SN~1994I.

In this work, we find that He features are persistent in all of our models, despite having only up to $\sim$10\% mass fraction of He in the outer layers, including the one that only contains 0.02~\Msun of He in the ejecta ($M_{\text{ZAMS}}$ = 45~\Msun). 
Lower ejecta mass (higher initial mass) models, such as the ones with $M_{\text{ZAMS}}$ = 60, 80, 100,  and 120 \Msun, show strong He features in both optical and NIR regions at all time ranges that are explored in this work (see Fig.~\ref{fig:tardis_spec_time_series}). 
While the high ejecta mass model ($M_{\text{ZAMS}}$ = 45~\Msun) shows strong He features only in NIR at early times and in both optical and NIR towards the maximum light.
At early time, \ion{He}{1} is still present in the optical region in this model but its contribution is weak and is heavily blended with other ions, as shown in the top left panel of Fig.~\ref{fig:TARDIS_SDEC_plot}.
After one month past explosions, the optical He feature strength quickly grows to the state that is compatible with other models that have $\sim$five times more He in their ejecta, such as the $M_{\text{ZAMS}}$ = 120~\Msun model.

The strong presence of the \ion{He}{1} 1.083~$\mu$m feature among all models at all times is consistent with the case study in \citet{Teffs2020}, which concludes that even trace amount of He can produce saturated NIR He lines. 
Although in our model spectra, the \ion{He}{1} 1.083~$\mu$m feature can be assessed to confirm the presence of He, both models and observations have shown that SESNe, including He-poor SNe-Ic objects, have a broad and strong absorption feature presenting around 1~$\mu$m that can be contributed by other elements such as \ion{C}{1}, \ion{O}{1}, and \ion{Mg}{2} \citep[\eg,][]{Dessart2015, Williamson2021, Shahbandeh2022}.

The fact that the models explored in this work consistently show strong He feature(s) does not exclude the possibility that a larger amount can be hidden in other types of progenitors and/or with different computational treatments of the helium. 
To further constrain the hidden He problem, a larger grid of progenitor models should be explored more thoroughly with consistent computational assumptions. 
Furthermore, the He feature is also sensitive to the \Nifs mass and mixing degree, a fixed input in this work due to the lack of nuclear-burning networks in the explosion stage.

We note that the \texttt{recomb-NLTE} approximation for He treatment in TARDIS is limited to cases when \ion{He}{2} ground state is the dominant state among the He population \citep{Boyle2017}.
This approximation was developed to test the double detonation scenario in SNe~Ia models by \citet{Boyle2017} and verified for SNe~Ic models by \citet{Williamson2021}.
In this work, we expand the application range to models containing non-neglectable amounts of He that spectroscopically may be identified as SNe~Ib.
In comparison with the SNe~Ib model from \citet{Hachinger2012}, the \texttt{recomb-NLTE} approximation produces consistent spectral line profiles in optical regions but overestimates the line strength of the He 2.058~$\mu$m feature, which is a better diagnosis for the presence of He in ejecta and more sensitive to the He mass compared to other lines in optical and NIR \citep{Dessart2015, Williamson2021, Shahbandeh2022}.
The two metastable \ion{He}{1} excited states, which are responsible for producing NIR \ion{He}{1} lines, are directly coupled with the \ion{He}{2} ground state and may be overestimated in conditions when a significant amount of He is recombined, especially in low-temperature regions.
Further improvement of the He treatment is needed for modeling SNe~Ib/c in the NIR region using \tardis.
Comparison with other radiative transfer codes with detailed balanced level population calculations and non-thermal effects is encouraged.

\section{Conclusion} \label{sec: conclusion}

We present synthetic light curves and spectra of physics-driven explosions of high-mass progenitor stars. 
The chosen progenitors are solar-metallicity and non-rotating single WR stars with $M_{\text{ZAMS}}$ range from 45 to 120~\Msun simulated using \kepler, taken from \citet{Sukhbold2016}.
Similar to \citet{Barker2022}, we map the physics-driven explosion outcomes,  simulated by \citet{Couch2020} using \flash with \stir mode, onto \snec to obtain the bolometric light curves. 
These explosions of these stripped stars are SESNe candidates with ejecta masses ranging from 4 to 11~\Msun.
We then calculate the synthetic spectra using \tardis based on the \snec outputs.

The model light curves of these high-initial-mass progenitor stars are broad and faint compared to the general population of observed SESNe \citep{Lyman2016, Taddia2016, Prentice2019, Rodriguez2023}.
The rise and decline rates of the model bolometric light curves are shown to be marginally consistent with observations of SESNe with broad light curves, which accounts for $\sim$15\% of the SESN population in the most recent study \citep{Karamehmetoglu2023}.
The rise time and peak luminosity of the model light curve are not only sensitive to the \Nifs mass and distribution but also rely on the opacity prescription in the numerical simulation (see H23).

Using the analytic formulas from \citet{Wheeler2015} and H23 for ejecta mass estimation, we found that the calibrated method in H23 provides the most consistent results within the calibrated ejecta mass range, with uncertainties below 20\%. 
However, for the model with the highest ejecta mass (10.85~\Msun), all methods exhibit large uncertainties, reaching up to 80\%.

We found that \ion{He}{1} lines are persistent in all our models in the NIR region, even in the one with only 0.02~\Msun of helium.
In the model with the least amount of helium in the ejecta, optical \ion{He}{1} lines are initially weak but quickly increase to the same strength as other models with larger helium mass.
The relatively low amount of helium required to show line profiles is consistent with the case studies carried out by \citet{Teffs2020} and \citet{Williamson2021}.
However, we caution that helium features strongly rely on the radiation field and composition, which requires further study to make quantitative claims. 
We note that the lack of a nuclear burning network is a major limitation of this work, as well as the assumption that \ion{He}{2} ground state is the dominating state in spectral simulation.

Our synthetic light curves and spectra are publically available\footnote{\Datalink}, and provide comparison resources for a large number of transients that will be detected by the upcoming large surveys, such as those carried out by the Vera C. Rubin Observatory and the Nancy Grace Roman Space Telescope.


\section*{ACKNOWLEDGEMENT}
The authors would like to thank the anonymous referee for their helpful comments.
We thank Chris Ashall, Anirban Dutta, Abigail Polin, Stefan Hachinger, and Melissa Shahbandeh for their helpful discussion. 
This research made use of \tardis, a community-developed software package for spectral synthesis in supernovae \citep{Kerzendorf2014, kerzendorf_2024_13207705}. 
The development of \tardis received support from GitHub, the Google Summer of Code initiative, and ESA's Summer of Code in Space program. 
\tardis is a fiscally sponsored project of NumFOCUS. 
\tardis makes extensive use of Astropy and Pyne.
This work was supported in part by Michigan State University through computational resources provided by the Institute for Cyber-Enabled Research.
J.L. is supported by NSF-2206523 and DOE No. DE-SC0017955.  
W.E.K. acknowledges financial support from NSF-2206523 and NSF-2311323.
M.M. and the METAL group at UVa acknowledge support in part from ADAP program grant No. 80NSSC22K0486, from the NSF grant AST-2206657 and from the HST GO program HST-GO-16656.
J.S. is supported by NSF OAC-2311323 and DOE No. DE-SC0017955
The Flatiron Institute is supported by the Simons Foundation.
This work was supported by the U.S. Department of Energy through the Los Alamos National Laboratory. Los Alamos National Laboratory is operated by Triad National Security, LLC, for the National Nuclear Security Administration of U.S. Department of Energy (Contract No. 89233218CNA000001).
BLB gratefully acknowledges support from the Los Alamos National Laboratory Metropolis Fellowship.
This article is cleared for unlimited release, LA-UR-24-30218.

\software{Astropy\footnote{\url{https://www.astropy.org/}} \citep{astropy:2013,astropy:2018, astropy:2022}, 
Pandas\footnote{\url{https://pandas.pydata.org/}} \citep{pandas_paper,pandas_software},
Scipy\footnote{\url{https://scipy.org/}} \citep{2020SciPy-NMeth},
SNEC\footnote{\url{https://stellarcollapse.org/SNEC.html}} \citep{Morozova2015, Morozova2016, Morozova2018},
TARDIS\footnote{\url{https://github.com/tardis-sn/tardis}} (\citealt{Kerzendorf2014}; version: \citealt{kerzendorf_2024_13207705})}

{\it Contributor Roles:} J.L. led the analysis and writing of the paper. W.E.K. led the project and oversaw the project progress. J.L. and B.L.B led the modeling efforts. J.G. and M.M. contributed to the analysis and interpretation. S.M.C., J.S., and A.G.F contributed relevant scientific expertise to the project.

\bibliographystyle{aasjournal}
\bibliography{SN_Ref,Softwares_ref}

\end{document}